\documentclass[11pt]{article} 
\pdfoutput=1 
\begin{document} 
\title{A falling droplet as it falls apart: \\ A Fluid Dynamics Video} 
\author{M. Jalaal$^1$, M. Schwalbach$^2$, K. Mehravaran$^1$ \\ 
\\\vspace{6pt} $^1$University of British Columbia, Kelowna, Canada.
\\ $^2$RWTH Aachen University, Aachen, Germany.} 
\maketitle 
\begin{abstract} 
Using direct numerical simulations, the fragmentation of falling liquid droplets 
in a quiescent media is studied. Three simulations with different E$\ddot{o}$tv$\ddot{o}$s 
numbers were performed. An adaptive volume of fluid (VOF) method based on octree meshing 
is used, providing a notable reduction of computational cost. The current video 
includes 4 main parts describing the fragmentation of the falling droplet.  
\end{abstract} 
\section{Part 1. Fragmentation of the droplet and vorticity changes in carrier phase
 } 
Temporal variation of the vorticity magnitude during the falling procedure is 
demonstrated here. A high vorticity region is obvious on top of the troplet. The more 
the fragments, the more the complexity in the carrier phase.    
 
\section{Part 2. Variation of falling velocity over the parent droplet 
and the fragments
 } 
Two of the crucial parameters for the case of falling droplets are size and velocity of 
the fragments. This video shows the variation of falling velocity over the interface.
Using the integration, the falling velocity and size of the fragments can be easily and 
accurately achieved.

\section{Part 3. Grid adaption
 } 
Since an accurate prediction of the radius of curvature and vorticity magnitude are essential
to the breakup process, these parameters are considered as refinement criteria, while the 
latter is also important to resolve the local turbulent eddies generated in the continues 
phase and close to the interface.
 
\section{Part 4. Mechanism of breakup
 } 
In this part, the detailed mechanism of thin liquid sheet breakup is shown where the instability 
growth over the interface generates several punctures. The retraction of punctures produces a
network of ligaments rapidly disintegrates into several number of small droplets. 

\section{Conclusion
 } 
A video presenting the results obtained via direct numerical simulation is presented. Outcomes
demonstrates the capability of the current numerical methods to investigate the complex problem
of falling droplets.

\end{document}